\documentclass[aps,prl,twocolumn,reprint,superscriptaddress]{revtex4-1}
\usepackage{graphicx}
\usepackage{amsmath}
\usepackage{amssymb}
\usepackage{colordvi}
\usepackage{mathrsfs}
\usepackage{bm}
\usepackage{verbatim}
\usepackage{dcolumn}
\usepackage{epsfig}
\usepackage{subfigure}
\usepackage[colorlinks,allcolors=blue]{hyperref}
\usepackage{ulem}
\usepackage{dsfont}
\usepackage{makecell}

\makeatother
\begin{document}
\title{Disorder-Induced Phase Transitions in Three-Dimensional Chiral Second-Order Topological Insulator}
\author{Yedi Shen}
\affiliation{International Centre for Quantum Design of Functional Materials, CAS Key Laboratory of Strongly-Coupled Quantum Matter Physics, and Department of Physics, University of Science and Technology of China, Hefei, Anhui 230026, China}
\author{Zeyu Li}
\affiliation{International Centre for Quantum Design of Functional Materials, CAS Key Laboratory of Strongly-Coupled Quantum Matter Physics, and Department of Physics, University of Science and Technology of China, Hefei, Anhui 230026, China}
\author{Qian Niu}
\email[Correspondence author:~~]{niuqian@ustc.edu.cn}
\affiliation{International Centre for Quantum Design of Functional Materials, CAS Key Laboratory of Strongly-Coupled Quantum Matter Physics, and Department of Physics, University of Science and Technology of China, Hefei, Anhui 230026, China}
\author{Zhenhua Qiao}
\email[Correspondence author:~~]{qiao@ustc.edu.cn}
\affiliation{International Centre for Quantum Design of Functional Materials, CAS Key Laboratory of Strongly-Coupled Quantum Matter Physics, and Department of Physics, University of Science and Technology of China, Hefei, Anhui 230026, China}
\affiliation{Hefei National Laboratory, University of Science and Technology of China, Hefei 230088, China}
\date{\today}

\begin{abstract}
Topological insulators have been extended to higher-order versions that possess topological hinge or corner states in lower dimensions. However, their robustness against disorders is still unclear. Here, we theoretically investigate the phase transitions of three-dimensional (3D) chiral second-order topological insulator (SOTI) in the presence of disorders. Our results show that, by increasing disorder strength, the nonzero densities of states of side surface and bulk emerge at disorder strengths of $W_{S}$ and $W_{B}$, respectively. The spectral function indicates that the bulk gap is only closed at one of the $R_{4z}\mathcal{T}$-invariant points, i.e., $\Gamma_{3}$. The closing of side surface gap or bulk gap is ascribed to the significant decrease of the elastic mean free time of quasi-particles. {To precisely confirm the phase boundaries for phase transitions, based on the scaling theory of localization length, we obtain the fixed points $W_{C 1}$ and $W_{C 2}$ as two critical disorder strengths, where the 3D chiral SOTI goes into a diffusive metallic phase and an Anderson insulating phase, respectively.}
\end{abstract}

\maketitle

\textit{Introduction}.---Higher-order topological insulators, characterized by hinge or corner states protected by various spatiotemporal symmetries{~\cite{PhysRevLett.98.106803,PhysRevB.76.045302,PhysRevB.103.085408,PhysRevB.99.085406,PhysRevB.96.245115,PhysRevLett.119.246402,PhysRevLett.108.126807,PhysRevLett.110.046404,PhysRevB.97.155305,PhysRevB.99.235125,PhysRevLett.123.186401,PhysRevB.98.241103,PhysRevB.98.245102,PhysRevB.98.081110,PhysRevB.98.201114,PhysRevB.98.205129,s41567-018-0224-7,PhysRevLett.125.037001,PhysRevLett.127.026803,PhysRevB.104.134508,EPL.142.56002}}, have invigorated many research fields, such as spintronics and phononics~\cite{10.1126/sciadv.aat0346,PhysRevLett.123.247401,PhysRevResearch.3.033177,Shiryu,Heqiu Li,phononics1,phononics2}. Although these states have been extensively observed in bosonic systems~\cite{bosons1,bosons2}, the observations are extremely limited in electronic systems. In particular, the three-dimensional (3D) chiral second-order topological insulator (SOTI), possessing gapped bulk states, gapped side surface states, and one-dimensional topologically-protected in-gap hinge states propagating unidirectionally, has not yet been experimentally observed. Given the ubiquitous disorders in crystalline materials, it is crucial to understand their robustness against disorders{~\cite{PhysRevResearch.2.033521,PhysRevB.103.115118,PhysRevLett.126.206404,PhysRevResearch.2.043197}}. In addition, without spin-orbit coupling and magnetic field, a disorder-induced metal-insulator transition exists in 3D electron systems, but does not exist in one and two dimensions~\cite{PhysRevLett.42.673}. Therefore, 3D chiral SOTIs and lower-dimensional topological insulators may exhibit significantly different behavior under disorders.

Based on renormalization-group calculations, it was reported that 3D chiral SOTIs are always unstable against Coulomb interaction and disorders~\cite{PhysRevLett.127.176601}, which has attracted widespread discussion~\cite{PhysRevB.106.155116,arXiv:2202.01642,arXiv:2202.03417}. However, some key information was missing. For example, they just considered the disorder-induced one-loop self-energy correction, and only one of the $R_{4z}\mathcal{T}$-invariant (combination of fourfold rotation and time-reversal symmetry) $\mathbf{k}$ points ($\Gamma_{1}$) were used to characterize the phase transition. Here, we reexamine their robustness against disorders by considering all the $R_{4z}\mathcal{T}$-invariant $\mathbf{k}$ points. We find that bulk gap closes at $\Gamma_{3}$, i.e., $(k_{x},k_{y},k_{z})=(0,0,\pi)$, but not $\Gamma_{1}$ as in previous reports. Meanwhile, by considering multiple scattering events~\cite{PhysRevB.92.085410,PhysRevLett.117.056802,PhysRevB.100.054108,PhysRevB.101.205424,PhysRevB.103.214201} which is beyond the abilities of the self-consistent Born approximation~\cite{PhysRevResearch.2.033521,PhysRevB.103.115118,PhysRevB.65.245420,PhysRevLett.125.166801,PhysRevB.74.235443} and the renormalization-group approach~\cite{PhysRevLett.127.176601,PhysRevB.74.235443}, we find the renormalized parameters can not be used as the unique criterion for a phase transition under disorders. It is necessary to consider the broadening of the energy spectrum caused by multiple scattering events.

\begin{figure}[t]
  \includegraphics[width=8.8cm]{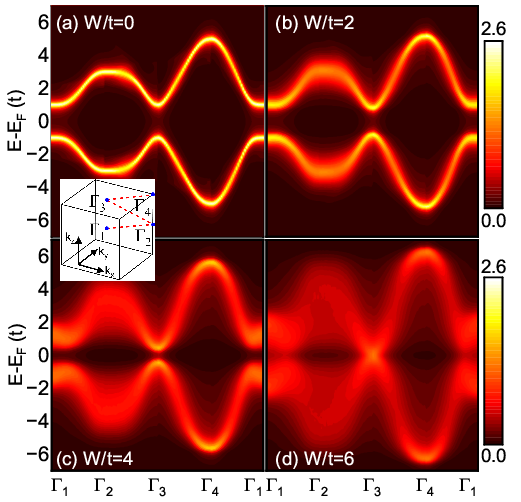}
  \caption{(Color online). The quasi-particle spectral function $A(\mathbf{k},E)$ along the high symmetry line consisting of four $R_{4 z} \mathcal{T}$-invariant $\mathbf{k}$ points. The color plot is drawn on a logarithmic scale. We adopt a large sample $L^{3}=160^{3} a^{3}$. (a)-(d) The quasi-particle spectral function of disordered 3D chiral SOTI with $W/t=0$, $2$, $4$, and $6$. Inset: Brillouin zone of the 3D chiral SOTI. $\Gamma_{1}$, $\Gamma_{2}$, $\Gamma_{3}$, and $\Gamma_{4}$ are $R_{4 z} \mathcal{T}$-invariant $\mathbf{k}$ points.}
	\label{fig1}
\end{figure}

In this Letter, we systematically study the phase transitions of the 3D chiral SOTI in the presence of random scalar disorders. By investigating the density of states and the averaged inverse participation ratio, we find that the side surface gap and bulk gap successively close at disorder strengths of $W_{S}$ and $W_{B}$, respectively. {The phase transitions are expected to take place around the two disorder strengths.} Based on the accurate momentum-space Lanczos method~\cite{PhysRevB.82.153405,PhysRevB.85.073407,PhysRevLett.118.146401}, that can rigorously treat all multi-scattering events from disorders, we obtain scaling properties of low energy quasi-particles in disordered 3D chiral SOTI around all four $R_{4z} \mathcal{T}$-invariant $\mathbf{k}$ points in the first Brillouin zone. Surprisingly, the spectral function and self-energy show that the four $R_{4 z}\mathcal{T}$-invariant $\mathbf{k}$ points exhibit different behaviors with the increase of disorder strength [see Fig.~\ref{fig1}], i.e., random scalar disorders only close the local gap at $\Gamma_{3}$. {Utilizing the scaling theory of localization length~\cite{PhysRevLett.47.1546,PhysRevLett.56.R7053}, we provide a phase diagram to show the phase boundaries of disorder-induced phase transitions.} {Because of the localization of side surface states, we find that the 3D chiral SOTI is robust until disorder strength reaches another critical strength $W_{C 1}$, where the 3D chiral SOTI goes into a diffusive metallic phase.} {And then, once the disorder strength exceeds a critical strength $W_{C2}$}, only located bulk states exist around the Fermi level, indicating the system further transiting from a diffusive metallic phase to an Anderson insulating phase.

\textit{Model for 3D chiral SOTI}---The tight-binding model on a simple cubic lattice can model a 3D chiral SOTI~\cite{10.1126/sciadv.aat0346}
\begin{equation}
	\begin{aligned}
		H_{0} & =\frac{M}{2} \sum_{\mathbf{r}, \alpha}(-1)^\alpha C_{\mathbf{r}, \alpha}^{\dagger} \sigma_0 C_{\mathbf{r}, \alpha} \\
		&+\frac{1}{2} \sum_{\mathbf{r}, \alpha} \sum_{i=x, y, z} t_{i}(-1)^\alpha C_{\mathbf{r}+\hat{e}_i, \alpha}^{\dagger} \sigma_0 C_{\mathbf{r}, \alpha}\\
		&+\frac{\Delta_1}{2 i} \sum_{\mathbf{r}, \alpha} \sum_{i=x, y, z} C_{\mathbf{r}+\hat{e}_i, \alpha+1}^{\dagger} \sigma_i C_{\mathbf{r}, \alpha} \\
		& +\frac{\Delta_2}{2 i} \sum_{\mathbf{r}, \alpha} \sum_{i=x, y, z}(-1)^\alpha n_i C_{\mathbf{r}+\hat{e}_i, \alpha+1}^{\dagger} \sigma_0 C_{\mathbf{r}, \alpha}+\text { h.c. },
		\label{chiral_HOTI}
	\end{aligned}
\end{equation}
where $M$ is the mass term. $t_{i}$, $\Delta_1$, and $\Delta_2$ are the nearest-neighbor hopping parameters. $\alpha$  ($0$ or $1$) represents the orbital subspace, $\hat{n}=(1,-1,0)$, and $C_{\mathbf{r}, \alpha}^{\dagger}(C_{\mathbf{r}, \alpha})$ is the creation (annihilation) operator with spin ($s=\uparrow$ or $\downarrow$) at the lattice site $\mathbf{r}$. $\sigma_0$ and $\sigma_i$ $(i=x, y, z)$ are $2\times2$ identity matrix and Pauli matrices, respectively, for the spin degree of freedom. The basis vectors are spanned by $\hat{e}_i$ $(i=x, y, z$). For the Hamiltonian~\ref{chiral_HOTI}, a nonzero $\Delta_2$ term stands for orbital currents breaking of both time-reversal symmetry ($\mathcal{T}$) and fourfold rotation symmetry ($R_{4z} \equiv \tau_0 e^{-i(\pi / 4) \sigma_z}$) at the same time. When $1<\left|M/t \right|<3$, the Hamiltonian~\ref{chiral_HOTI} falls into the region of a 3D chiral SOTI. For simplicity, we set $t_{i}=-t=-1$, $M/t=2$, $\Delta_{1}/t=1$, and $\Delta_{2}/t=1$, ensuring the same system topology~\cite{PhysRevLett.127.026803}.

\begin{figure*}[t!]
  \includegraphics[width=18.cm]{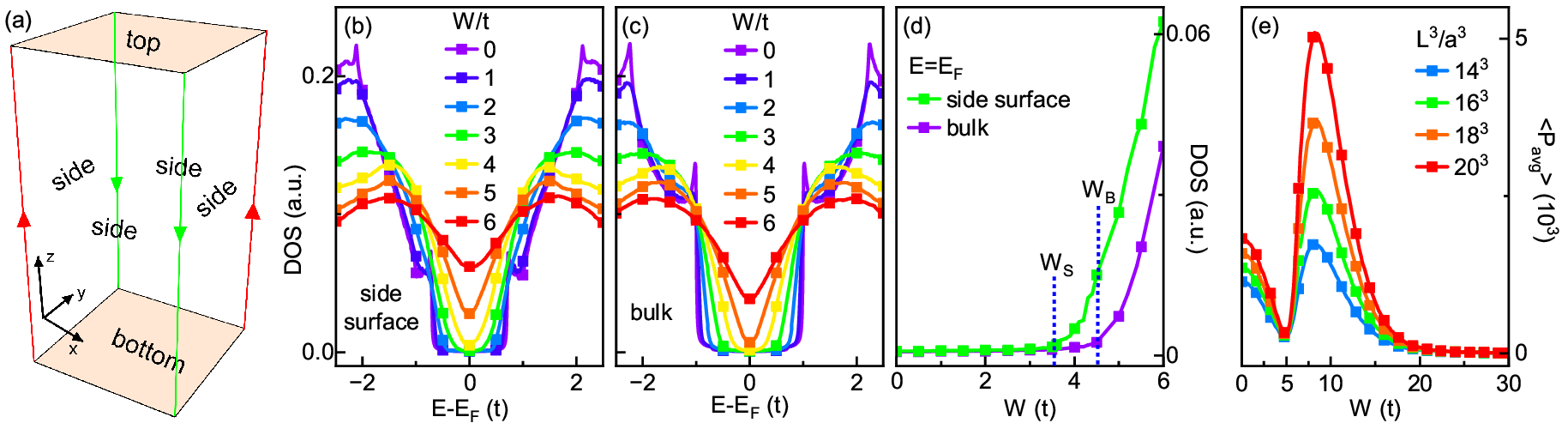}
  \caption{(Color online). (a) Schematic of a 3D chiral SOTI with open boundary conditions. (b)-(c) In a large sample of $L^3=160^3 a^3$, the side surface and bulk density of states for a disordered 3D chiral SOTI with $W/t=0,1,2,3,4,5,$ and $6$. (d) The density of states at $E=E_{F}$ vary with disorder strength. With the increase of disorder strength, the widths of the side surface gap and the bulk gap gradually decrease, and finally close at $W_{S}/t \approx 3.5$ and $W_{B}/t \approx 4.5$, respectively. {(e) The averaged inverse participation ratio as a function of disorder strength $W$ for different sample size $L^{3}/a^{3}=14^{3}, 16^{3}, 18^{3},$ and $20^{3}$ with open boundary conditions at $E=E_{F}$. Over 1000 ensembles are collected for each point.}}
  \label{fig2}
\end{figure*}

\textit{Density of states and spectral function.}---First, we study the density of states of a 3D chiral SOTI in the presence of random scalar disorders, which is included as $V_{\text {dis}}= V(\mathbf{r}) I_{4\times 4}$. $V(\mathbf{r})$ is uniformly distributed between $-W/2$ and $W/2$, where $W$ represents the disorder strength. And the $R_{4z}\mathcal{T}$ symmetry is preserved under disorders~\cite{PhysRevLett.127.176601}. For a 3D chiral SOTI, the chiral hinge states are located in the bulk and side surface gaps. {Hence, the occurrence of a phase transition will be sensitive to both the magnitude of the bulk gap and the side surface gap.} Therefore, the evolution of the local density of states of the side surface and bulk can be used to characterize the robustness of hinge states against disorders. The local density of states can be evaluated as $\rho\left(\mathbf{r}_i, E\right)=-\operatorname{Im} \langle i |\frac{1}{E-H+i \eta} | i \rangle / \pi$. Here, a small artificial broadening parameter of $\eta=0.01t$ is employed to simulate an infinitesimal imaginary energy. Based on the well-developed Lanczos recursive method~\cite{PhysRevB.77.195411,PhysRevLett.102.056803}, it is possible to numerically calculate an accurate local density of states. The bulk or side surface density of states can be obtained by taking either the mean of all local density of states or the ensemble average. To attain a high energy resolution and reduce finite-size errors, a large sample $\left(L^3=160^3 a^3\right)$ with open boundary conditions in three dimensions is considered [see Fig.~\ref{fig2}(a)]. Figure~\ref{fig2}(b) displays the bulk density of states as a function of energy $E-E_{F} (t)$ for different disorder strengths, i.e., $W/t=0$, 1, 2, 3, 4, 5, and 6. A pristine 3D chiral SOTI has a wide bulk gap determined by $\Delta_{1}$. With the increase of $W$, the side surface gap gradually decreases and remains open until the disorder strength reaches $W_{S}/t\approx3.5$. Once the disorder strength exceeds $W_{S}$, a nonzero side surface density of states emerges at the Fermi level $E_{F}$ implying the close of the side surface gap. Moreover, the bulk density of states exhibits a similar behavior as a function of energy $E-E_{F} (t)$ for different disorder strengths [see Fig.~\ref{fig2}(c)]. Because the bulk gap is larger than the side surface gap, and it closes at a stronger disorder strength of $W_{B}/t\approx4.5$. {The phase transitions are expected to take place around the two disorder strengths.} The bulk and side surface density of states at $E_{F}$ as a function of $W(t)$ are displayed in Fig.~\ref{fig2}(d).

{As mentioned above, the closings of side surface gap and bulk gap signify the phase transitions in the presence of disorders. To confirm the existence of new phases, we elaborate on the  averaged inverse participation ratio expressed as~\cite{PhysRevLett.115.076601,PhysRevLett.127.236402}
\begin{equation}
	P_{\mathrm{avg}}=\left\langle\frac{\left[\sum_{i, \alpha,s}\left|\psi_{\alpha,s}\left(\mathbf{r}_{i}\right)\right|^{2}\right]^{2}}{\sum_{i, \alpha,s}\left|\psi_{\alpha,s}\left(\mathbf{r}_{i}\right)\right|^{4}}\right\rangle,
\end{equation}
where the wave function $\psi_{\alpha,s}\left(\mathbf{r}_{i}\right)$ is calculated at site $i$ with orbit $\alpha$, spin $s$, and $E=E_{F}$. $\langle\cdots\rangle$ denotes the disorders average. As a example, we plot the averaged inverse participation ratio as a function of $W$ at $E=E_F$ for different volumes $L^{3}/a^{3}=14^3, 16^3, 18^3$, and $20^3$ [see Fig.~\ref{fig2}(e)]. It is well known that the averaged inverse participation ratio scales as $P_{\mathrm{avg}}\sim L^{d}$ in a $d$-dimensional metallic phase and $P_{\mathrm{avg}}\sim \text{const}$ in an insulating phase. In the weak disorders region, $P_{\mathrm{avg}} \sim L$, which implies that there are one-dimensional metallic chiral hinge states. With the increase of $W$, the extensibility of the metallic chiral hinge states gradually becomes worse. After that, $P_{\mathrm{avg}} \sim L^3$, which implies that the massive extended bulk states have a primary impact on the averaged inverse participation ratio, i.e., the 3D chiral SOTI goes into a diffusive metallic phase. For even larger disorder strength, $P_{\mathrm{avg}} \sim \text{const}$, corresponding to an Anderson insulating phase. It is worth noting that there is no a 3D first-order topological insulator present here with a large averaged inverse participation ratio and $P_{\mathrm{avg}} \sim L^2$.}

To further illustrate the effects of disorders, the properties of quasi-particle in momentum space are studied. We model a 3D chiral SOTI system with a low-energy effective Bloch Hamiltonian~\cite{10.1126/sciadv.aat0346,PhysRevLett.127.176601}
\begin{equation}
	\begin{aligned}
		H_0(\mathbf{k})= & {\left[M+\sum_i t_{i}  \cos \left(a k_i\right)\right] \tau_z \sigma_0+\Delta_1\sum_i  \sin \left(a k_i\right) } \\
		& \times \tau_x \sigma_i+\Delta_2\left[\cos \left(a k_x\right)-\cos \left(a k_y\right)\right] \tau_y \sigma_0,
	\end{aligned}
	\label{Eq:imaginaryselfenergy}
\end{equation}
where $\sigma_i$ and $\tau_i$ ($i = x,y,z$) are the Pauli matrices for spin and orbital degrees of freedom, respectively. $a$ is the lattice constant. The combination of $R_{4z}$ and $\mathcal{T}$ is preserved and generates four $R_{4z}\mathcal{T}$-invariant $\mathbf{k}$ points at $\Gamma_i$, where $\Gamma_i$ belongs to $\{(0,0,0),(\pi, \pi, 0),(0,0, \pi),(\pi, \pi, \pi)\}$ ($i=1,2,3,4$), respectively. For the valence and conduction bands, the eigenvalues of $H_{0}(\mathbf{k})$ are $E_{0}(\Gamma_{1})=\pm 1$, $E_{0}(\Gamma_{2})=\pm 3$, $E_{0}(\Gamma_{3})=\pm 1$, and $E_{0}(\Gamma_{4})=\pm 5$, where $\pm$ represents different orbitals. Each band has two spins, resulting in a two-fold degeneracy. In a large 3D sample with millions of atoms $\left(L^3=160^3 a^3\right)$, we analyze the modification of the energy spectra in momentum space based on the accurate momentum-space Lanczos recursive method~\cite{PhysRevB.82.153405,PhysRevB.85.073407,PhysRevLett.118.146401}, which can capture all multi-scattering events. The quasi-particle spectral function is bridged with the Green's function through the equation $A(\mathbf{k}, E)=-\operatorname{Im} G(\mathbf{k}, E) / \pi$~\cite{PhysRevB.85.073407}. The energy spectrum calculated along the high-symmetry line, which consists of four $R_{4z}\mathcal{T}$-invariant $\mathbf{k}$ points, is displayed in Fig.~\ref{fig1}. When $W/t=0$, the spectral function $A_0(\mathbf{k}, E)$ is a $\delta$ function, suggesting that the wave vector $\mathbf{k}$ is a good quantum number and all its weight is concentrated at the energy $E=E_{\mathbf{k}}$ [see Fig.~\ref{fig1}(a)]. In the presence of disorders, the $\delta$ peak becomes broadened due to the disorder-scattering effect, giving a finite elastic mean free time to quasi-particle, and the bulk gap begins to gradually decrease [see Figs.~\ref{fig1}(a)-(b)]. The peak width is determined by the imaginary part of the self-energy, $\operatorname{Im} \Sigma(E)$. After entering a strong scattering region, the spread of the spectral function becomes prominent, and the bulk gap eventually closes at $\Gamma_{3}$, while the others remain open, as shown in Fig.~\ref{fig1}(d). We also find $\left| \langle \Gamma_{3},\alpha,s|V_{\text {dis}}|\Gamma_{3},\alpha,s \rangle \right|\gg\left| \langle \Gamma_{1},\alpha,s |V_{\text {dis}}|\Gamma_{3},\alpha,s \rangle \right|$, which means intra-valley scattering at {$\Gamma_{3}$} is stronger than inter-valley scattering from $\Gamma_{1}$ to $\Gamma_{3}$. So, we can only focus on the $\Gamma_{3}$ to research the disorder-induced phase transition in momentum space.

\begin{figure}[t!]
	\includegraphics[width=1.0\linewidth]{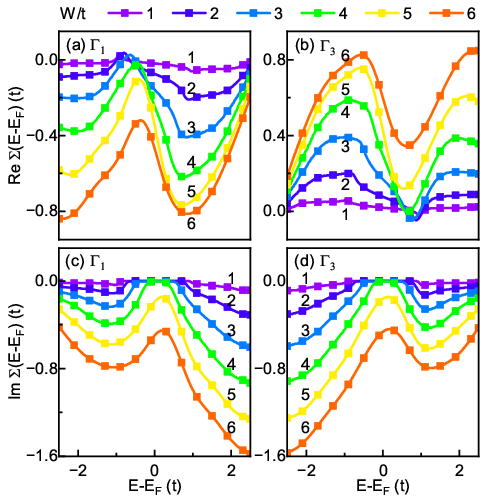}
	\caption{(Color online). (a)(b) Real and (c)(d) imaginary parts of the self-energy as a function of energy for different disorder strengths ($1 \leq W/t \leq 6$) of the valence band at $\Gamma_{1}$ and $\Gamma_{3}$, respectively. At $\Gamma_{1}$/$\Gamma_{3}$, with the increase of disorder strength, the conduction and valence bands move away/closer from/to the Fermi level.}
	\label{fig3}
\end{figure}

\textit{Accurate self-energy of a disordered 3D chiral SOTI.}---By utilizing the accurate momentum-space Lanczos recursive method, the phase transition can be further understood through the accurate self-energy solved by the Dyson equation: $\Sigma(\mathbf{k}, E)=G_0^{-1}(\mathbf{k}, E)-G^{-1}(\mathbf{k}, E)$. Figure~\ref{fig3}(a) plots the real part of the quasi-particle self-energy of the valence band at $\Gamma_{1}$ for different disorder strengths. Due to disorder effects, the roots of $E-E_{F}-E_{0}(\mathbf{k})=\operatorname{Re} \Sigma(E-E_{F})$ correspond to the quasi-particle dispersion $E_{\mathbf{k}}$, implying a decrease in the energy of quasi-particles. Furthermore, the elastic mean free time is inversely proportional to the imaginary part of the self-energy, given by $\tau=\left[\hbar /-2  \operatorname{Im} \Sigma\left(E-E_{F}\right)\right]$, which can be used to describe the decay time of quasi-particles. As shown in Fig.~\ref{fig3}(c), the elastic mean free time at $\Gamma_{1}$ gradually decreases with the increase of disorder strength. Because of the particle-hole symmetry, the self-energy obtained based on the eigenstate basis satisfies the following relations: $\operatorname{Re} \Sigma(\Gamma_{1}, E,\alpha_{i},s)=-\operatorname{Re} \Sigma(\Gamma_{1},-E,\alpha_{j},s)$ and $\operatorname{Im} \Sigma(\Gamma_{1}, E,\alpha_{i},s)=\operatorname{Im} \Sigma(\Gamma_{1},-E,\alpha_{j},s)$ ($i\neq j$), which ensure that the dispersion relations are symmetric with respect to the Fermi level. And we find that $\operatorname{Re} \Sigma(\Gamma_{1}, E, \alpha_{i},s)=-\operatorname{Re} \Sigma(\Gamma_{3},-E, \alpha_{j},s)$ and $\operatorname{Im} \Sigma(\Gamma_{1}, E, \alpha_{i},s)=\operatorname{Im} \Sigma(\Gamma_{3},-E, \alpha_{j},s)$ ($i=j$), as shown in Fig.~\ref{fig3}(b) and~\ref{fig3}(d). From a similar analysis, we find that the energy of quasi-particle at $\Gamma_ {3}$ increases, and the elastic mean free time decreases with the increase of disorder strength.

Furthermore, we also explore the correction to the hopping parameters induced by disorders. By doing a unitary transformation, we transform the self-energy from the eigenstate basis to the orbital-spin basis. Then, we can construct an effective Hamiltonian including disorders, i.e., $H_{\text{eff}}=H_{0}(\mathbf{k})+U_{\mathbf{k}}\Sigma(E)U_{\mathbf{k}}^{\dagger}$. Due to multiple scattering events, the small elastic mean free time effectively broadens the spectral function. Therefore the corrected $\Delta_{1}$ and $\Delta_{2}$ are non-vanishing, the bulk gap and side surface gap become closed. As functions of the running scale parameter, the renormalized $\Delta_1$ and $\Delta_2$ go to zero~\cite{PhysRevLett.127.176601}, but they can not be used as a unique criterion for the phase transition in the presence of random scalar disorders. The broadening of the energy spectrum eventually leads to the closing of the bulk gap.

\begin{figure}[t!]
	\includegraphics[width=1.0\linewidth]{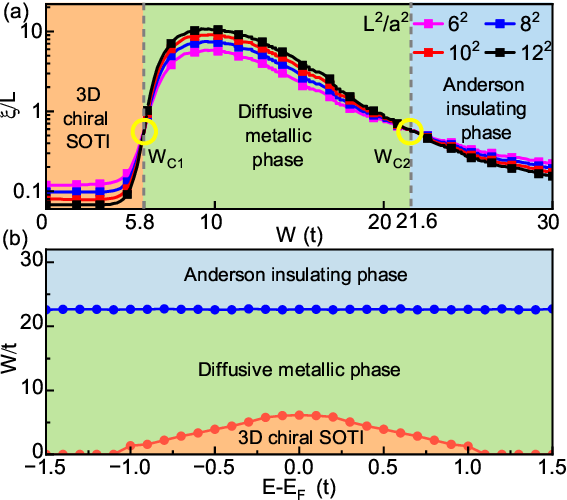}
	\caption{(Color online). {(a) Normalized localization length $\xi / L$ as a function of the disorder strength $W$ at $E=E_{F}$ calculated on quasi-two-dimensional slices, with a length of $1 \times 10^6$ and different areas of $L^{2}/a^{2}=6^2, 8^2, 10^2$ and $12^2$. $W_{C 1}/t \approx 5.8$ and $W_{C 2}/t \approx 21.6$ are two critical points.  (b) Phase diagram in $(E-E_F, W)$ plane.}}
	\label{fig4}
\end{figure}

\textit{Phase transitions.}---{As mentioned above, three phases will successively emerge with the increase of disorder strength, i.e., a 3D chiral SOTI, a diffusive metallic phase and an Anderson insulating phase. To precisely confirm the phase boundaries, we numerically calculate the localization length $\xi$ on a quasi-two-dimensional slice of essentially infinite length $\left(1 \times 10^6\right)$ and finite area $L^{2}$ by using the transfer-matrix method~\cite{PhysRevLett.47.1546,PhysRevLett.56.R7053}. The periodic condition is applied to eliminate the possible hinge-state transport. We plot the normalized localization length $\xi /L$ as a function of $W$ at $E=E_F$ for different areas $L^{2}/a^{2}=6^2, 8^2, 10^2$, and $12^2$ [see Fig.~\ref{fig4}(a)]. One can find that there are two fixed points, $W_{C 1}/t \approx 5.8$ and $W_{C 2}/t\approx21.6$. When $W<W_{C 1}$, $\xi / L$ decreases with the increase of $L^2$, indicating that $\xi / L$ will converge to zero when $L^2 \rightarrow \infty$, signaling a localized insulating phase, i.e., the 3D chiral SOTI. When $W_{C 1}< W < W_{C 2}$, $\xi / L$ increases with the increase of $L$, indicating that $\xi / L$ will diverge when $L^2 \rightarrow \infty$, signaling a delocalized metallic phase. When $W>W_{C 2}$, $\xi / L$ behaves similarly to that in the weak disorders case, meaning that it enters an Anderson insulating phase. Therefore, the fixed points $W_{C 1}$ and $W_{C 2}$ are two critical disorder strengths for the insulator-metal and metal-insulator phase transitions, respectively. To build the phase diagram in the $(E-E_{F}, W)$ plane, we calculate the critical disorder strengths $W_{C 1}$ and $W_{C 2}$ for different energies, which define the phase boundaries [see Fig.~\ref{fig4}(b)]. }

\textit{Conclusion.}---Based on accurate numerical calculation methods, we systematically analyze the disorder-driven phase transitions of the 3D chiral SOTI in the presence of random scalar disorders. The density of states and spectral function indicate that the side surface gap and bulk gap successively close at disorder strengths of $W_{S}/t=3.5$ and $W_{B}/t=4.5$, respectively. It is noted that the bulk gap is only closed at one of the $R_{4z}\mathcal{T}$-invariant $\mathbf{k}$ points, i.e., $\Gamma_{3}$. We also obtain the accurate self-energy to build an effective Hamiltonian, revealing that the close of the bulk gap ascribes to the reduced elastic mean free time of quasi-particles, which leads to a broadening of the spectral function. {When the disorder strength is beyond $W_{C1}$ and $W_{C2}$, the 3D chiral SOTI can be successively driven into two different phases: a diffusive metallic phase, and an Anderson insulating phase, respectively.} Our results provide a clear picture to distinguish the disorder-driven phase transitions of the 3D chiral SOTI.

\begin{acknowledgments}
This work was financially supported by the National Natural Science Foundation of China (Grant Nos. 11974327 and 12004369), the Fundamental Research Funds for the Central Universities (Nos. WK3510000010 and WK2030020032), Innovation Program for Quantum Science and Technology (2021ZD0302800), and Anhui Initiative in Quantum Information Technologies (Grant No. AHY170000). We also thank for the high-performance supercomputing services provided by the Supercomputing Center of University of Science and Technology of China.
\end{acknowledgments}

\end{document}